\documentstyle[12pt,fleqn]{article}
\def\soc{{\rm C}_{60}}
\def\rug{{\rm C}_{70}}
\def\c76{{\rm C}_{76}}
\def\c78{{\rm C}_{78}}
\def\beeq{\begin{equation}}
\def\eneq{\end{equation}}
\def\beeqa{\begin{eqnarray}}
\def\eneqa{\end{eqnarray}}

\addtocounter{section}{-1}
\begin{document}

\begin{center}

\vspace{2in}

{\large {\bf{Optical absorption spectra of A$_{\bf 6}$C$_{\bf 60}$:\\
Reduction of effective Coulomb interactions
} } }

\vspace{1cm}

{\rm Kikuo H{\sc arigaya}\footnote[1]{E-mail address: harigaya@etl.go.jp.}}\\

\vspace{1cm}

{\sl Fundamental Physics Section,\\
Electrotechnical Laboratory,\\
Umezono 1-1-4, Tsukuba, Ibaraki 305}

\vspace{1cm}

(Received~~~~~~~~~~~~~~~~~~~~~~~~~~~~~~~~~~~)
\end{center}

\Roman{table}

\vspace{1cm}

\noindent
{\bf Abstract}\\
Optical absorption spectra of $\soc^{6-}$ are theoretically
investigated in order to analyze the optical properties of
alkali metal doped fullerites A$_6\soc$.  We use a tight binding
model with long ranged Coulomb interactions and bond disorder.
Optical spectra are obtained by the Hartree-Fock approximation
and the CI method.  The Coulomb interaction parameters which
are relevant to the optical spectra of A$_6\soc$ are almost the
half of those of the neutral $\soc$.  The reduction of the effective
Coulomb interactions is concluded for the heavily doped case of $\soc$.

\mbox{}


\pagebreak

\section{Introduction}

Since the $\soc$ solid and A$_3\soc$$^{1)}$ with the novel
high temperature superconductivity were discovered, fullerenes
have been intensively investigated.  As the $\pi$ electrons
are delocalized on their surfaces, the fullerenes show optical
responses that are similar to those in $\pi$ conjugated polymers.$^{2)}$
For example, the absorption spectra of $\soc$ (refs. 3 and 4)
and $\rug$ (ref. 3) reflect the existence of excitons
(mainly Frenkel excitons) which are important when the excitation
energy is larger than the order of 1eV.  The nonlinearity of $\soc$
in the third harmonic generation (THG) is of the order
$10^{-11}$esu,$^{5,6)}$ and the similar magnitudes have been observed
in polydiacetylenes.

Recently, we have studied the linear absorption and the THG of
$\soc$ by using a tight binding model$^{7)}$ and a model
with a long ranged Coulomb interaction.$^{8,9)}$  A free electron model
yields the THG magnitudes which are in agreement with the
experiment of $\soc$.$^{5,6)}$  However, when the Coulomb interactions
are taken into account, the THG magnitudes decrease.$^{8)}$  We
have discussed that the local field correction would be necessary
in order to recover the agreement.  The model with Coulomb interactions
has been turned out to describe well the linear absorption spectra of
$\soc$ and $\rug$ in solutions.$^{9)}$

In this paper, we consider the optical spectra of A$_6\soc$ (A = K, Rb,
Cs, etc).  The system is a fullerite maximumly
doped with alkali metals, and is an
insulator like the neutral $\soc$.  The optical spectra have been
measured in several papers.$^{10,11)}$  The peak structures in the
energy dependence are quite different from those of the neutral $\soc$,
even though each peak could be explained as an optical transition
between molecular orbitals.  The absorption spectra of the neutral
$\soc$ and A$_3\soc$ are rather similar, but the data of A$_6\soc$
are largely different.  It seems that main peaks of the neutral
$\soc$ spectra move to lower energies.  This fact could be explained
by reduction of the effective Coulomb interaction strengths.
The principal purpose of the present calculations is to confirm this
conjecture.  The calculation method is the same as that used in the
previous paper.$^{9)}$  We start from the Hartree-Fock approximation
and perform configuration interaction calculations which are limited
to single electron-hole excitations (single CI).  In the experiments of
A$_6\soc$,$^{10,11)}$ intermolecular interactions seem to be relatively
weak, because there is not a peak structure owing to the intermolecular
interactions (like the 2.8eV structure in the $\soc$ films$^{4)}$).
Therefore, we shall limit our considerations to the system $\soc^{6-}$.

In the next section, we explain our model briefly.  In \S 3, we
show results and give discussion relating with molecular orbital
structures.  The final section is devoted to the summary.

\section{Model}

In order to consider optical spectra of $\soc^{6-}$,
we use the following hamiltonian:
\beeq
H = H_0 + H_{\rm bond} + H_{\rm int}.
\eneq
The first term $H_0$ of eq. (1) is the free electron part:
\beeq
H_0 = -t \sum_{\langle i,j \rangle, \sigma}
( c_{i,\sigma}^\dagger c_{j,\sigma} + {\rm h.c.} ),
\eneq
where $c_{i,\sigma}$ is an anihilation operator of the $\pi$-electron at
site $i$ with spin $\sigma$; the sum is taken over all the pairs
$\langle i,j \rangle$ of neighboring atoms; and $t$ is the hopping integral.
In the $\soc^{6-}$ which is geometrically optimized by the SSH model,$^{12)}$
the dimerization is negligible.  So, we assume the constant hopping integral.
The second term of eq. (1) is the bond disorder model simulating
lattice fluctuations:
\beeq
H_{\rm bond} = \sum_{\langle i,j \rangle, \sigma} \delta t_{i,j}
(c_{i,\sigma}^\dagger c_{j,\sigma} + {\rm h.c.}).
\eneq
This model was sometimes used in the literatures.$^{9,13)}$
The third term is the long ranged Coulomb interaction in the form
of the Ohno potential:
\beeqa
H_{\rm int} &=& U \sum_i
(c_{i,\uparrow}^\dagger c_{i,\uparrow} - \frac{1}{2})
(c_{i,\downarrow}^\dagger c_{i,\downarrow} - \frac{1}{2})\\ \nonumber
&+& \sum_{i \neq j} W(r_{i,j})
(\sum_\sigma c_{i,\sigma}^\dagger c_{i,\sigma} - 1)
(\sum_\tau c_{j,\tau}^\dagger c_{j,\tau} - 1),
\eneqa
where $r_{i,j}$ is the distance between the $i$th and $j$th sites and
\beeq
W(r) = \frac{1}{\sqrt{(1/U)^2 + (r/r_0 V)^2}}
\eneq
is the Ohno potential.  The quantity $W(0) = U$ is the strength of
the onsite interaction, and $V$ means the strength of the long range part.

In this paper, most of the quantities with the energy dimension are shown
in the units of $t$.  The same convention has been used in the previous
paper.$^{9)}$  We vary the Coulomb interaction strengths within
$0 \leq V \leq U \leq 5t$ in order to fix appropriate parameters.
When disorder effects are included, data are averaged over 100 samples.
This is enough to obtain smooth numerical data.

\section{Results and Discussion}

Figure 1(a) shows the optical spectrum of the free electron model.
Four low energy peaks are named with Roman symbols.  The meaning
of each transition is interpreted in the energy level structures
shown as Fig. 2.  The $t_{1u}$ orbital near the energy 0 is filled
up with six electons, and the orbitals above it are not occupied.
The $\soc^{6-}$ has the closed shell structure.  The absorption
spectrum of the neutral $\soc$ has been calculated in ref. 7.
The peaks, b and d, are present in that calculation.  The oscillator
strengths are the same as in the present calculation.
The peaks, a and c, are characteristic to $\soc^{6-}$.
The experiments$^{10,11)}$ have interpreted the lowest
feature around 1.2eV as originated from the peak a.  It seems that
this assignment is a natural understanding.

In the previous study,$^{9)}$ we have shown that the peak positions
and the distribution of oscillator strengths in the optical absorption
of the neutral $\soc$ are reasonably explained by the theory with
$U = 4t$, $V = 2t$, and $t = 1.8$eV.  This is the intramolecular
excitonic effect.  We shall apply the same formalism to $\soc^{6-}$.
If the same parameters, $U = 4t$, $V = 2t$, and $t = 1.8$eV, are relevant
to $\soc^{6-}$, we theoretically expect  that the peak at 2.0$t$ does
not move so much, as the doping proceeds.  However, in experiments,$^{9)}$
the corresponding 3.6eV peak of the neutral $\soc$ moves to 2.8eV
in $\soc^{6-}$.  This large shift cannot be explained by the theory
with the same parameters, because the half of characters of optical
transitions are the same: the transition b shown in Fig. 2 is common
to $\soc$ and $\soc^{6-}$.

We have searched for Coulomb parameters which reproduce overall features
of the optical spectrum of K$_6$$\soc$ (ref. 10).  We find that the
parameters, $U = 2t$, $V = 1t$, and $t=2.0$eV, would be reasonable.
Figure 1(b) shows the calculated spectrum.  The several peaks in
Fig. 1(a) move to higher energies, and oscillator strengths become
relatively larger in higher energies.  These are the Coulomb
interaction effects.

Experimental spectra$^{10,11)}$ are broad mainly due to lattice
fluctuations.  The effects are simulated by the bond disorder model with
Gaussian distribution of the standard deviation $t_{\rm s} = 0.20t$.
The result is shown in Fig. 1(c).  The experimental data are taken from
ref. 10, and are shown as Fig. 1(d) for comparison.
We could say that two features around
1.2eV and 2.8eV are reasonably explained by the present calculation.
The broadening is simulated well by the bond disorder.  The disorder
strength $t_{\rm s} = 2.0t$ is about the twice
as large as that of the neutral $\soc$.
This indicates that A$_6\soc$ is more disordered than the neutral $\soc$.
The broad feature in the energies higher than 3.6eV in the experiment
might correspond to the two broad peaks around 2.2$t$ and 3.0$t$ of
the calculation.  In these energies, the agreement is not so
good.  The same fact has been seen in the neutral $\soc$ case.$^{9)}$
Excitations which include $\sigma$ orbitals would be mixed in this
energy region.  This effect could be taken into account by using
models with $\pi$ and $\sigma$ electrons.  However, this is beyond
our interests.

Our main conclusion, the reduction of the effective Coulomb interactions,
seems to be the remarkable difference between the neutral and heavily
doped $\soc$.  The electron density of $\soc^{6-}$ is 10\% larger than that
of the neutral $\soc$.  The relevant Coulomb interaction strengths could
be different in heavily doped molecules and solids.  One of the origins
of the reduction would be the fact that the environments around $\soc$
molecules are different between the neutral $\soc$ and A$_6\soc$.  Actually,
the $\soc$ solid is the simple cubic lattice,
and A$_6\soc$ is the body centered cubic lattice.  The distance
between the neighboring $\soc$ is larger in
A$_6\soc$ due to the intercalated alkali metal ions.  The research of the
intermolecular interaction effects in the neutral $\soc$ crystal is
in progress, and will be reported elsewhere.$^{14)}$  We should consider
interactions among molecules in A$_6\soc$, and calculate the optical
spectra in future.

\section{Summary}

We have considered optical excitations of $\soc^{6-}$ by treating
the long ranged Coulomb interaction with the Hartree-Fock approximation
and the single CI method.  We have searched for the parameter
set in order to reproduce overall features
of the optical spectrum of A$_6\soc$.  The parameters, $U \sim 2t$,
$V \sim 1t$, and $t \sim 2.0$eV, have been turned out to be relevant.
The magnitudes of $U$ and $V$ are almost the half of those which were used
for the neutral $\soc$.  This might be due to the different environments
around molecules in crystals.  Calculations which include intermolecular
interactions would be desirable for further understandings of electronic
structures.

\mbox{}

\noindent
{\bf Acknowledgements}\\
The author acknowledges useful discussion with Dr. Shuji Abe.
He is grateful for helpful correspondences with Prof. Bret C. Hess
and Dr. Mitsutaka Fujita.

\pagebreak
\begin{flushleft}
{\bf References}
\end{flushleft}

\noindent
1) A. F. Hebard, M. J. Rosseinsky, R. C. Haddon, D. W. Murphy,
S. H. Glarum, T. T. M. Palstra, A. P. Ramirez, and A. R. Kotran:
Nature {\bf 350} (1991) 121.\\
2) A. J. Heeger, S. Kivelson, J.R. Schrieffer, and W. P. Su:
Rev. Mod. Phys. {\bf 60} (1988) 781.\\
3) J. P. Hare, H. W. Kroto, and R. Taylor: Chem. Phys. Lett.
{\bf 177} (1991) 394.\\
4) S. L. Ren, Y. Wang, A. M. Rao, E. McRae, J. M. Holden, T. Hager,
KaiAn Wang, W. T. Lee, H. F. Ni, J. Selegue, and P. C. Eklund:
Appl. Phys. Lett. {\bf 59} (1991) 2678.\\
5) J. S. Meth, H. Vanherzeele, and Y. Wang: Chem. Phys. Lett.
{\bf 197} (1992) 26.\\
6) Z. H. Kafafi, J. R. Lindle, R. G. S. Pong, F. J. Bartoli,
L. J. Lingg, and J. Milliken: Chem. Phys. Lett. {\bf 188}
(1992) 492.\\
7) K. Harigaya and S. Abe: Jpn. J. Appl. Phys. {\bf 31}
(1992) L887.\\
8) K. Harigaya and S. Abe: J. Lumin. (to be published).\\
9) K. Harigaya and S. Abe: Phys. Rev. B (to be published).\\
10) T. Pichler, M. Matus, J. K\"{u}rti, and H. Kuzmany:
Solid State Commun. {\bf 81} (1992) 859.\\
11) W. L. Wilson, A. F. Hebard, L. R. Narasimhan, and
R. C. Haddon: Phys. Rev. B {\bf 48} (1993) 2738.\\
12) K. Harigaya: Phys. Rev. B {\bf 48} (1993) 2765.\\
13) S. Abe, M. Schreiber, W. P. Su, and J. Lu: Mol. Cryst. Liq. Cryst.
{\bf 217} (1992) 1.\\
14) S. Abe and K. Harigaya: (in preparation).\\

\pagebreak

\begin{flushleft}
{\bf Figure Captions}
\end{flushleft}

\mbox{}

\noindent
Fig. 1.  Optical absorption spectra for $\soc^{6-}$ shown in
arbitrary units.  The abscissa is scaled by $t$.
The spectra are calculated with the parameters (a) $U = V = 0$,
(b,c) $U = 2t$ and $V = 1t$.  The Lorentzian broadening
$\gamma = 0.01 t$ is used.  In (c), lattice fluctuations
are taken into account by the bond disorder of the strength
$t_{\rm s} = 0.20t$.  (d) The experimental spectrum (ref. 10)
of K$_6\soc$ is shown with using $t = 2.0$eV.

\mbox{}

\noindent
Fig. 2.  Energy level structures of the free electron (H\"{u}ckel) model.
Each energy level is shown against its degeneracy.  Several dipole allowed
transitions are indicated by the Roman symbols.  The corresponding peak
positions in Fig. 1(a) are shown by the same symbols.

\end{document}